\documentclass{nst}

\usepackage{subfigure,dcolumn}
\usepackage[T2A,T1]{fontenc}
\usepackage[russian,english]{babel}

\usepackage{listings}
\begin{document}

\title{Efficient production of $^{229m,g}$Th via neutron capture in VUV-transparent crystals}\thanks{Supported by Fundamental and Interdisciplinary Disciplines Breakthrough Plan of the Ministry of Education of China-JYB2025XDXIM204, the National Natural Science Foundation of China (NSFC) (Grant Nos. 11974043, Nos. 11921006, Nos. 12405282, and Nos. 12305266) and the National Grand Instrument Project (No.2019YFF01014400).}

\author{Zhong-yi Chen}
\affiliation{Department of Physics, School of Mathematics and Physics, University of Science and Technology Beijing, Beijing 100083, China}
\affiliation{State Key Laboratory of Nuclear Physics and Technology, School of Physics, CAPT, Peking University, Beijing 100871, China}

\author{Hao-yang Lan}
\email[Corresponding author, ]{haoyang_lan@163.com.}
\affiliation{State Key Laboratory of Nuclear Physics and Technology, School of Physics, CAPT, Peking University, Beijing 100871, China}

\author{Di Wu}
\affiliation{State Key Laboratory of Nuclear Physics and Technology, School of Physics, CAPT, Peking University, Beijing 100871, China}

\author{Mei-zhi Wang}
\affiliation{State Key Laboratory of Nuclear Physics and Technology, School of Physics, CAPT, Peking University, Beijing 100871, China}

\author{Ze Chen}%
\affiliation{Department of Physics, School of Mathematics and Physics, University of Science and Technology Beijing, Beijing 100083, China}

\author{Li-pan Qin}%
\affiliation{Department of Physics, School of Mathematics and Physics, University of Science and Technology Beijing, Beijing 100083, China}

\author{Mei-qi Sun}%
\affiliation{Department of Physics, School of Mathematics and Physics, University of Science and Technology Beijing, Beijing 100083, China}

\author{Yu-peng Chen}%
\affiliation{Department of Physics, School of Mathematics and Physics, University of Science and Technology Beijing, Beijing 100083, China}

\author{Yan Tian}%
\affiliation{Department of Physics, School of Mathematics and Physics, University of Science and Technology Beijing, Beijing 100083, China}

\author{Jin Yan}%
\affiliation{Department of Physics, School of Mathematics and Physics, University of Science and Technology Beijing, Beijing 100083, China}

\author{Yan Wang}%
\affiliation{Department of Physics, School of Mathematics and Physics, University of Science and Technology Beijing, Beijing 100083, China}

\author{Xun-jie Ma}%
\affiliation{Department of Physics, School of Mathematics and Physics, University of Science and Technology Beijing, Beijing 100083, China}

\author{Xun Zhu}%
\affiliation{Department of Physics, School of Mathematics and Physics, University of Science and Technology Beijing, Beijing 100083, China}

\author{Yu-Meng Dong}%
\affiliation{Department of Physics, School of Mathematics and Physics, University of Science and Technology Beijing, Beijing 100083, China}

\author{Xin-Lu Xu}%
\affiliation{State Key Laboratory of Nuclear Physics and Technology, 
School of Physics, CAPT, Peking University, Beijing 100871, China}
\affiliation{Beijing Laser Acceleration Innovation Center, Beijing 101407, China}

\author{Xue-qing Yan}
\email[Corresponding author, ]{x.yan@pku.edu.cn}
\affiliation{State Key Laboratory of Nuclear Physics and Technology, School of Physics, CAPT, Peking University, Beijing 100871, China}
\affiliation{Beijing Laser Acceleration Innovation Center, Beijing 101407, China}

\author{Yun-liang Wang}
\email[Corresponding author, ]{ylwang@ustb.edu.cn.}
\affiliation{Department of Physics, School of Mathematics and Physics, University of Science and Technology Beijing, Beijing 100083, China}

\begin{abstract}
The low-lying isomeric state of $^{229m}$Th, owing to its unique nuclear energy structure, has been widely regarded as one of the most promising candidates for the development of a nuclear clock. However, the limited availability of suitable $^{229}$Th sources with sufficiently high activity remains a major challenge for experimental investigations of the $^{229m}$Th isomer. In this work, we present a comprehensive simulation study of a neutron-capture-based approach for the in-situ production of $^{229m,g}$Th by doping $^{228}$Ra into crystal hosts, where $^{229m,g}$Th is generated through neutron capture followed by a sequence of radioactive decays. Furthermore, we systematically investigate the background contributions associated with the three doped crystal hosts, namely CaF$_2$, SrF$_2$, and LiF, and evaluate their impact on the detection and identification of $^{229m}$Th. Under a neutron flux of $10^{15}\ \mathrm{n/cm^{2}/s}$ and a $^{228}$Ra doping concentration of $10^{19}\ \mathrm{cm^{-3}}$, this approach is capable of producing on the order of $10^{12}$ $^{229}$Th and $^{229m}$Th nuclei within only 1 s of irradiation. The simulation results demonstrate that signal-to-noise ratios as high as $10^5$ can be achieved for all three crystal hosts, indicating the feasibility of detecting and identifying the generated $^{229m}$Th. In addition, the influences of detector wavelength resolution and post-irradiation measurement time on the detectability of the $^{229m}$Th signal are systematically analyzed, and the corresponding optimal measurement conditions are identified. Furthermore, the spatial distribution of neutron-produced $^{229}$Th within the crystal is investigated, providing practical guidance for optimizing crystal geometry and illumination configuration in future continuous-wave VUV absorption spectroscopy experiments. These results demonstrate that the neutron-activation approach provides a promising alternative pathway for the production and detection of $^{229\mathrm{m,g}}$Th, which may facilitate future studies toward the realization of nuclear-clock-based technologies.
\end{abstract}

\keywords{nuclear clock, $^{229m}$Th production, neutron activation, doped crystals.}

\maketitle

\section{Introduction}
The realization of ultra-precise frequency standards is of fundamental importance for both applied and fundamental physics.\cite{bib:1,bib:2,bib:3,bib:4,bib:5,bib:6}
While state-of-the-art atomic clocks have achieved remarkable accuracy,\cite{bib:7,bib:8,bib:9} nuclear clocks based on nuclear transitions are expected to provide even higher stability owing to their reduced sensitivity to external perturbations.\cite{bib:10,bib:11,bib:12}
Among all known nuclei, thorium-229 ($^{229}$Th) is distinguished by the existence of an exceptionally low-lying isomeric state, $^{229m}$Th, whose excitation energy lies in the eV range,\cite{bib:13,bib:14,bib:15,bib:16} several orders of magnitude lower than that of conventional nuclear transitions.
Combined with a long radiative lifetime of up to $10^4$~s,\cite{bib:17,bib:18} the isomer is expected to possess an extremely narrow relative linewidth of ${\Delta E/E} \approx {10^{-20}}$.
These remarkable properties permit direct laser excitation of a nuclear transition and render $^{229m}$Th one of the most promising candidates for the development of next-generation nuclear clocks with unprecedented precision.

In recent years, considerable research efforts have been devoted to investigating the fundamental properties of $^{229m}$Th.\cite{bib:13,bib:14,bib:15,bib:19,bib:20,bib:21}
The excitation energy of the isomer has been constrained to the vacuum ultraviolet (VUV) region around 8~eV with increasing precision, and the de-excitation mechanisms, including internal conversion\cite{bib:16,bib:22,bib:23,bib:24,bib:25} and radiative decay,\cite{bib:26,bib:27,bib:28,bib:29} have been extensively investigated in both atomic and solid-state environments.
A particularly significant advance has been the observation of radiative decay of $^{229m}$Th in large-bandgap VUV-transparent crystals such as CaF$_2$ and MgF$_2$,\cite{bib:26,bib:27} which established that the isomer embedded in a crystalline host can decay via photon emission rather than internal conversion.
This breakthrough, together with the subsequent direct laser excitation of the $^{229}$Th nuclear transition in $^{229}$Th:CaF$_2$ and $^{229}$Th:LiSrAlF$_6$ crystals,\cite{bib:30,bib:31} has paved the way toward a solid-state nuclear clock in which a large ensemble of $^{229}$Th nuclei can be simultaneously interrogated in a compact platform.\cite{bib:11,bib:12} These experimental breakthroughs have been enabled by significant progress in the growth of high-quality $^{229}$Th-doped VUV-transparent crystals\cite{bib:32,bib:33} and have recently culminated in precision laser spectroscopy of the nuclear transition referenced to the $^{87}$Sr atomic clock\cite{bib:34} as well as the demonstration of clock frequency reproducibility.\cite{bib:35} In parallel, a growing body of experimental and theoretical work has revealed that the $^{229m}$Th isomer is sensitive to its local solid-state environment: laser-induced quenching,\cite{bib:36} temperature-dependent frequency shifts,\cite{bib:37} and host-dependent variations in radiative decay behavior\cite{bib:38} have been observed, while the roles of internal conversion\cite{bib:39} and host-specific electronic structure\cite{bib:40} in determining the isomer's properties in crystalline matrices have been theoretically elucidated.

Despite these advances, the development of a solid-state nuclear clock faces two interrelated challenges.
The first is the efficient and controllable population of $^{229m}$Th, whose weak transition strength imposes stringent requirements on both the excitation source and the detection system.
The second, and equally critical, is the limited availability of $^{229}$Th itself, which serves as the essential starting material for both isomer population and nuclear clock operation.
These two challenges are deeply coupled: progress in populating and characterizing $^{229m}$Th ultimately depends on access to sufficient quantities of $^{229}$Th-doped materials, while advances in $^{229}$Th material fabrication rely on feedback from $^{229m}$Th detection and spectroscopy.
Addressing either challenge in isolation is therefore insufficient for the realization of practical nuclear clock technologies.

Several approaches have been explored to populate the isomeric state.
Early studies mainly relied on nuclear decay processes, in which $^{229m}$Th is populated as a daughter product following the decay of a parent nucleus.
Notable examples include the $\alpha$ decay of $^{233}$U,\cite{bib:41} the $\beta$ decay of $^{229}$Ac,\cite{bib:42} and the electron-capture decay of $^{229}$Pa produced via the $^{232}$Th$(p,4n)$ reaction.\cite{bib:43}
More recently, photonuclear approaches have been investigated,\cite{bib:44} in which high-energy $\gamma$ rays induce reactions such as $^{232}$Th$(\gamma,3n)$ to directly produce $^{229m}$Th, or $^{232}$Th$(\gamma,2np)$ to generate $^{229}$Ac whose subsequent $\beta$ decay populates the isomeric state.
In parallel, direct excitation schemes that promote existing $^{229}$Th nuclei to the isomeric state have also been pursued.
Synchrotron-radiation experiments at the SPring-8 facility successfully excited $^{229}$Th nuclei to the second excited state at 29~keV,\cite{bib:20,bib:27} leading to efficient population of $^{229m}$Th in $^{229}$Th-doped CaF$_2$ crystals through cascaded decay.
More recently, direct VUV laser excitation of the nuclear transition has been demonstrated,\cite{bib:30,bib:31} facilitated by state-of-the-art VUV laser systems\cite{bib:45} and innovative $^{229}$Th doping techniques.
Alternative mechanisms mediated by electron--nucleus coupling, including the electron bridge process and electron-impact excitation, have also attracted considerable interest as possible routes for populating the isomeric state.\cite{bib:46,bib:47,bib:48,bib:49}

A common feature of all direct excitation schemes is their reliance on existing $^{229}$Th as the target material.
However, the global inventory of $^{229}$Th is severely limited, as it is primarily obtained from the $\alpha$ decay of $^{233}$U.\cite{bib:41}
The use of $^{233}$U is strictly regulated owing to its potential applications in nuclear technology, and the ongoing disposal of the $^{233}$U stockpile further threatens the long-term supply. To mitigate these constraints, alternative material platforms such as $^{229}$ThF$_4$ thin films\cite{bib:50} and $^{232}$Th-doped SrF$_2$ crystals\cite{bib:51} have been explored, the latter enabling \textit{in situ} conversion into $^{229}$Th-doped variants via photonuclear reactions\cite{bib:44} while bypassing the handling challenges associated with highly radioactive $^{229}$Th precursors.
Moreover, $^{229}$Th is also a critical precursor for the production of $^{225}$Ac used in targeted alpha cancer therapy, introducing competing demands that may prioritize medical applications over fundamental research.\cite{bib:44}
Even under optimistic assumptions, the accessible supply of $^{229}$Th remains orders of magnitude below what would be needed for large-scale nuclear clock development and systematic studies of $^{229m}$Th across diverse host materials.
This scarcity motivates the exploration of alternative production routes that do not depend on existing $^{229}$Th inventories, and that can simultaneously generate both the ground-state $^{229}$Th --- as a doping material for solid-state nuclear clocks --- and the isomeric state $^{229m}$Th for spectroscopy and detection studies.

In this work, we investigate a neutron-capture-based approach for the \textit{in-situ} production of both $^{229}$Th and $^{229m}$Th by doping $^{228}$Ra into VUV-transparent crystal hosts. In this approach, $^{228}$Ra serves as a precursor nucleus, undergoing the $(n,\gamma)$ reaction to produce $^{229}$Ra, which subsequently $\beta$ decays via $^{229}$Ac to $^{229}$Th, including its isomeric state.
The reaction pathway is illustrated in Fig.~1.
Notably, $^{228}$Ra can be obtained from decay chains originating from $^{232}$Th, which has a natural abundance of about $99.98\%$ and is widely available in natural ores.
This provides a potentially accessible and sustainable source for the production of $^{229\mathrm{m,g}}$Th, bypassing the regulatory and supply constraints associated with $^{233}$U-derived $^{229}$Th.

\begin{figure*}[!htb]
\includegraphics[width=0.9\hsize]{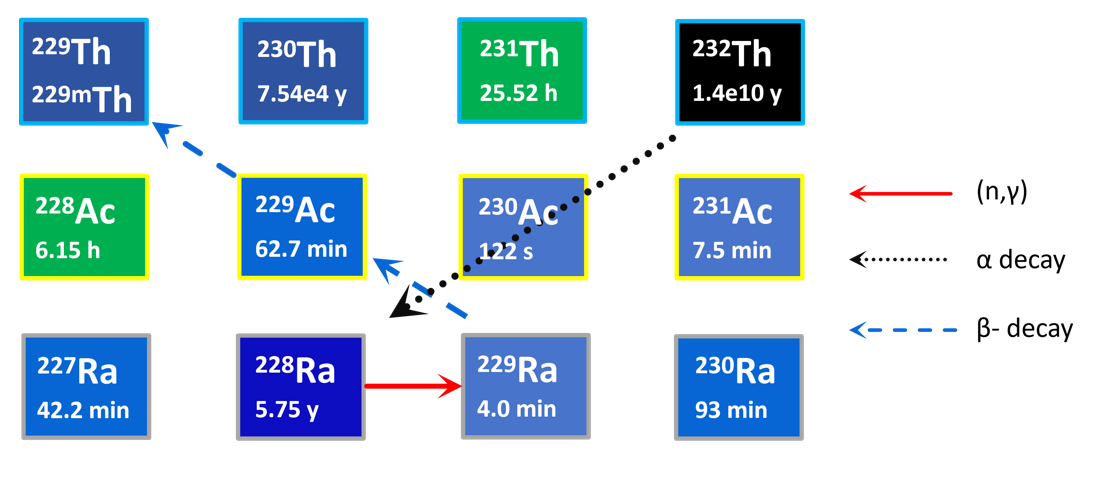}
\caption{Production pathway of $^{229}$Th and its isomer $^{229m}$Th via neutron capture on $^{228}$Ra and subsequent decay chain.}
\end{figure*}

While neutron activation of $^{228}$Ra-doped crystals has previously been suggested as a promising route for generating $^{229m}$Th,\cite{bib:52} a comprehensive evaluation of isotope production, radioactive backgrounds, signal detectability, and crystal optimization has not yet been carried out. A key advantage of the proposed approach is that $^{228}$Ra is doped directly into VUV-transparent crystals such as CaF$_2$, SrF$_2$, and LiF---all of which have been successfully demonstrated as host matrices for thorium doping with high optical quality\cite{bib:32,bib:33,bib:51}---which serve simultaneously as the production target, the detection medium, and --- after $^{229}$Th has been generated --- the host matrix for nuclear clock operation.
This integrated strategy eliminates the need for chemical separation and re-doping steps between isotope production and optical detection.
The in-situ generation of $^{229}$Th within the crystal volume also produces a spatial distribution of resonant nuclei that is inherently suited for continuous-wave (CW) VUV absorption spectroscopy,\cite{bib:53,bib:54,bib:55} a recently demonstrated technique that enables rapid and highly sensitive probing of the nuclear transition.
Thus, the method concurrently addresses both challenges identified above: it provides a scalable route for producing $^{229}$Th-doped crystals for solid-state nuclear clocks, while simultaneously generating $^{229m}$Th that can be detected and characterized through its radiative decay in the same crystal host.

In this work, we present a comprehensive simulation study of the neutron-activation approach in $^{228}$Ra-doped CaF$_2$, SrF$_2$, and LiF crystals using Geant4. We systematically investigate the production yields of $^{229}$Th and $^{229m}$Th under neutron irradiation, evaluate the Cherenkov and intrinsic radioactive backgrounds, and assess the detectability of the $^{229m}$Th radiative decay through signal-to-noise ratio (SNR) calculations under realistic detector conditions. The influences of detector wavelength resolution and post-irradiation measurement time are further analyzed to identify optimal detection conditions. Finally, the spatial distribution of neutron-produced $^{229}$Th within the crystal is investigated to provide practical guidance for optimizing crystal geometry and illumination configuration in future CW VUV absorption spectroscopy experiments.

The remainder of this paper is organized as follows.
Section~II describes the simulation framework, including the neutron energy spectrum, the $^{228}$Ra$(n,\gamma)$ reaction channel, the crystal doping conditions, and the resulting production yields of $^{229}$Th and $^{229m}$Th.
Section~III investigates the influence of Cherenkov radiation induced by charged particles from radioactive decay chains on the detection of $^{229m}$Th.
Section~IV evaluates the SNR for $^{229m}$Th detection, taking into account Cherenkov background contributions from both intrinsic $^{228}$Ra dopants and irradiation-induced radioactive products, as well as the effects of detector wavelength resolution and post-irradiation measurement time.
Section~V discusses the spatial distribution of $^{229}$Th production in crystals after neutron irradiation and its implications for CW VUV absorption spectroscopy.
Section~VI summarizes the main conclusions of this work.

\section{Neutron-Capture Reaction Cross-section and $^{229,229m}$Th Yields}
In this work, we assume a neutron flux of $10^{15}$ $n/cm^{2}/s$ for studying the $^{228}$Ra$(n,\gamma)$ reaction, with the corresponding neutron-capture cross section shown in Fig. 2. Such flux levels are currently achievable at several facilities, including the High Flux Isotope Reactor (HFIR) at Oak Ridge National Laboratory (ORNL),\cite{bib:56} the Advanced Test Reactor (ATR) at Idaho National Laboratory (INL),\cite{bib:57,bib:58} the China Advanced Research Reactor (CARR) at the China Institute of Atomic Energy (CIAE),\cite{bib:59} and the SM-3 reactor at the Research Institute of Atomic Reactors (RIAR) in Dimitrovgrad.\cite{bib:60} Therefore, the neutron irradiation conditions adopted in this work are representative of existing high-flux reactor facilities. The process was simulated using the Geant4 toolkit with the G4HadronPhysicsFTFP$\_$BERT$\_$HP physics list. The number density of purified $^{228}$Ra is $10^{19}$ cm$^{-3}$, doped in CaF$_2$, SrF$_2$, LiF crystal. Each crystal is modeled as a cylinder with a radius of 0.5 cm and a height of 1 cm. When the three radium-doped crystals were irradiated by a neutron source, various neutron reaction residuals with atomic numbers ranging from 2 to 90 were produced, as shown in Fig. 3. The residual nuclei are primarily distributed across three regions. The two low-Z regions arise from neutron-capture reactions on the crystal’s constituent elements and any subsequent radioactive decays, whereas the high-Z region results from neutron capture on $^{228}$Ac followed by its decay chain, leading to the production of $^{229}$Th and its isomer $^{229m}$Th. The production yields of $^{229}$Th and $^{229m}$Th are summarized in Table 1. The yields are generally comparable among the three radium-doped crystals and reach the order of $10^{12}$ nuclei per crystal after only 1 s of irradiation with a high-flux neutron source. These results highlight the advantage of the proposed method in achieving high production yields of both $^{229}$Th and $^{229m}$Th within a short irradiation time. However, the production of $^{229m}$Th in the LiF crystal is relatively lower than that in CaF$_2$ and SrF$_2$. Therefore, CaF$_2$ and SrF$_2$ may be preferable host materials for radium doping in applications targeting the production and detection of $^{229m}$Th. Furthermore, provided that the depletion of $^{228}$Ra is negligible and sufficiently high doping concentrations or sufficient crystal quantities are available, extending the irradiation time can further increase the production yields of both $^{229}$Th and $^{229m}$Th approximately proportionally.

\begin{figure}[!htb]
\includegraphics[width=\hsize]{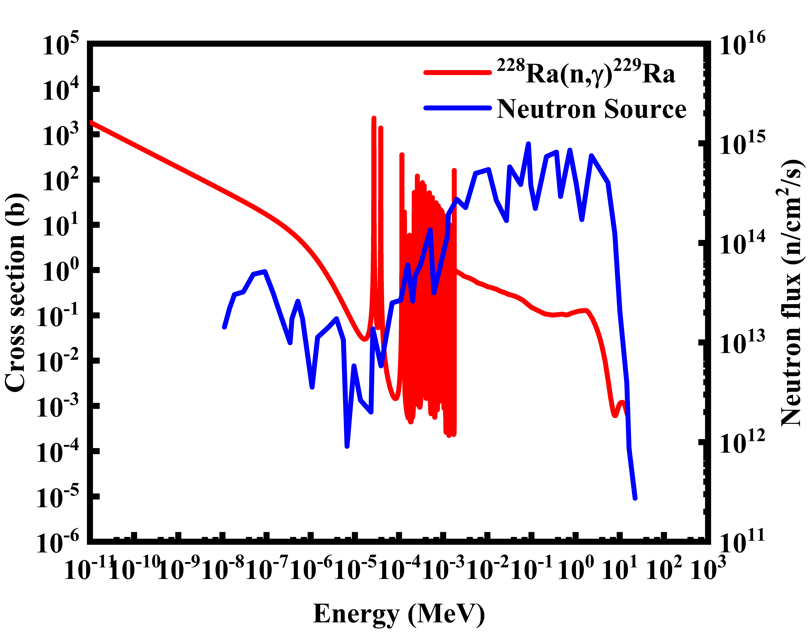}
\caption{Thermal neutron spectrum \cite{bib:61} and neutron-capture cross section of $^{228}$Ra.}
\end{figure}

\begin{figure}[!htb]
\includegraphics[width=\hsize]{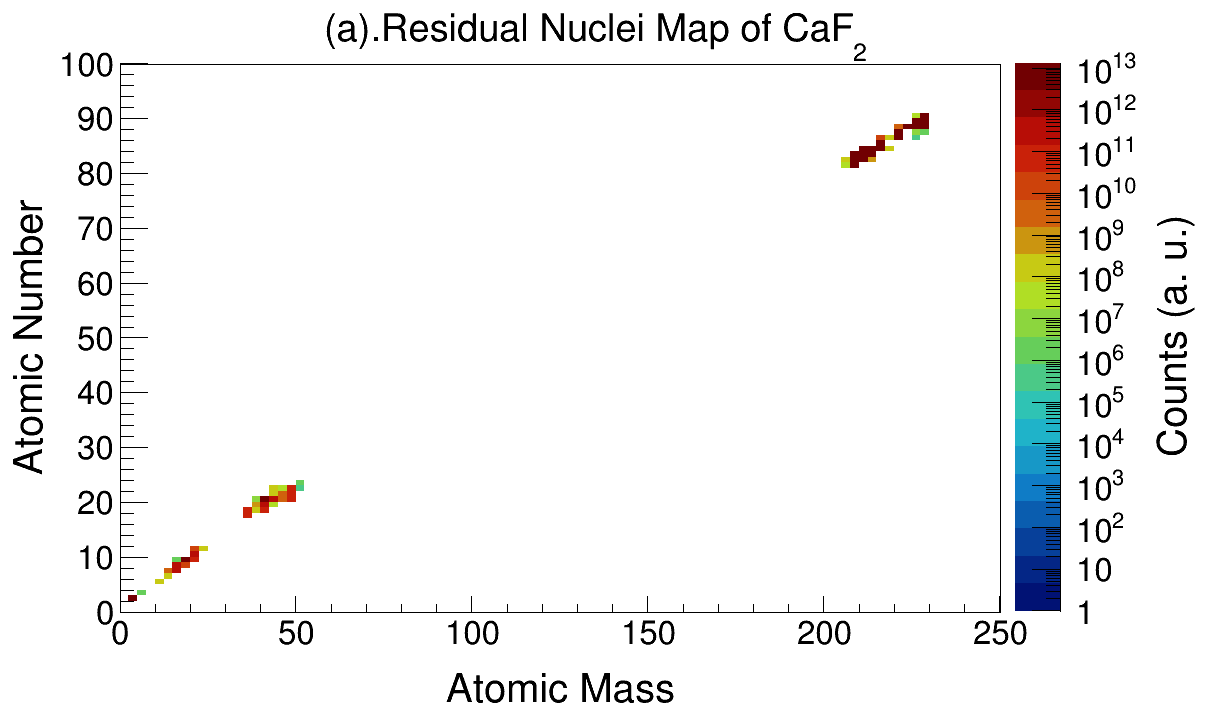}
\includegraphics[width=\hsize]{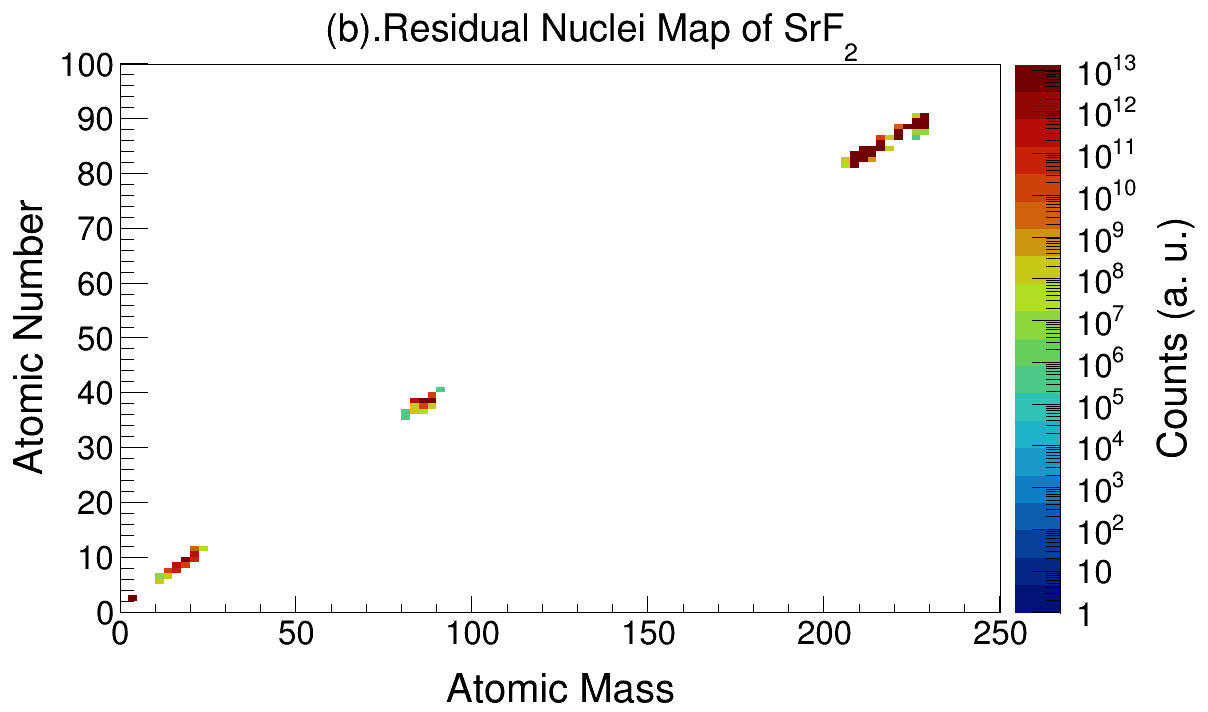}
\includegraphics[width=\hsize]{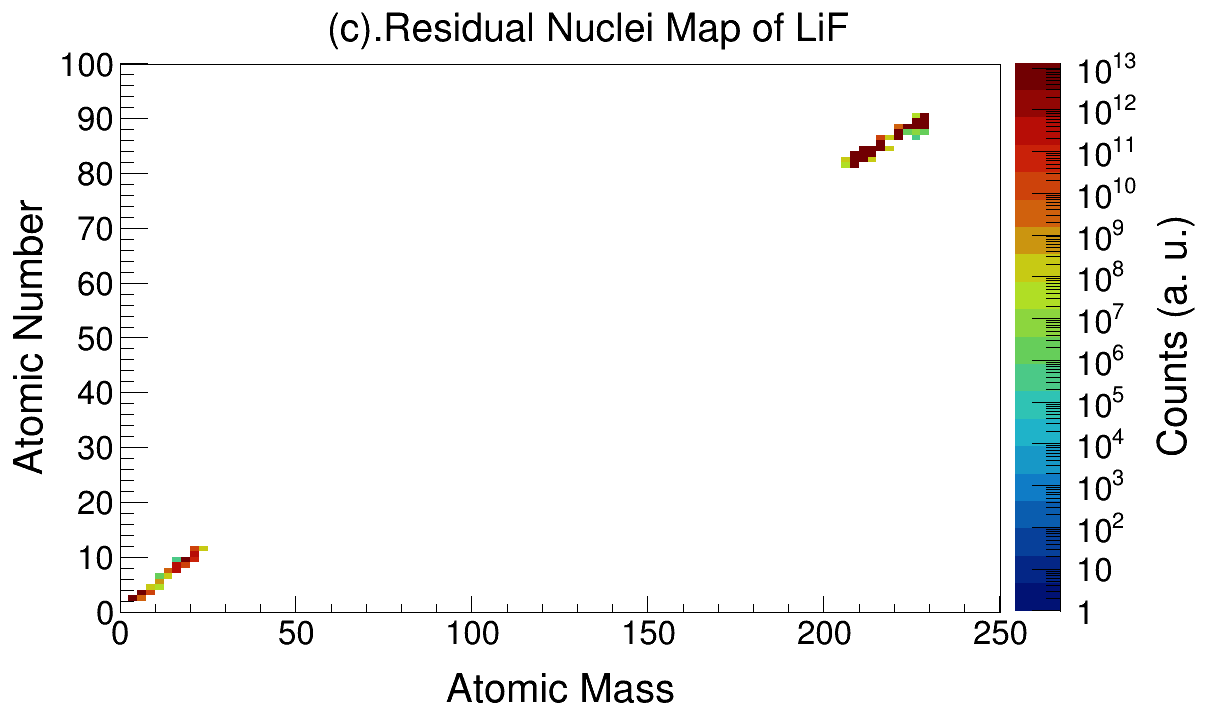}
\caption{The A-Z distribution of neutron reaction residuals produced in $^{228}$Ra doped in CaF$_2$(a), SrF$_2$(b), LiF(c) crystal.}
\end{figure}

\begin{table}[!htb]
\caption{The production yields of $^{229}$Th and $^{229m}$Th for the three radium-doped crystals.}
\begin{tabular*}{8cm}{@{\extracolsep{\fill}} lcc}
\toprule
\multicolumn{1}{c}{Doped Crystal} & \multicolumn{2}{c}{Production yield (/cm$^{3}$)} \\
\cmidrule(lr){2-3}
 & $^{229}$Th yield & $^{229m}$Th yield \\
\midrule
CaF$_2$  & $5.66\times10^{12}$ & $1.15\times10^{12}$ \\
SrF$_2$  & $6.01\times10^{12}$ & $1.22\times10^{12}$ \\
LiF      & $4.46\times10^{12}$ & $9.08\times10^{11}$ \\

\bottomrule
\end{tabular*}
\end{table}

\section{Cherenkov Background and VUV Signal Emission in Doped Crystals}
In the experiment, the generation of $^{229m}$Th in the crystal was inferred from the detection of its decay signals. Upon irradiation of the radium-doped crystals with a neutron source, the reaction products are radioactive and undergo decay, emitting electrons, positrons, $\gamma$ rays, or $\alpha$ particles. Therefore, it is necessary to evaluate whether these particles could affect the measurement of $^{229m}$Th. Taking radium-doped CaF$_2$ crystal as an example, we analyze the potential impact of these particles. The energy-time distribution of the emitted $\gamma$ rays is illustrated in Fig. 4(a). The red marker represents the $\gamma$-ray signal associated with the decay of $^{229m}$Th (with a transition energy of approximately 8.3 eV) in the $^{228}$Ra-doped CaF$_2$ crystal, with a half-life of about $10^{3.6}$ s, corresponding to the decay of $^{229}$Ac. Considering only the $\gamma$-ray signal, no interfering signals are observed in the vicinity of the $^{229m}$Th decay peak. Therefore, with a high-resolution detector, this signal should be clearly distinguishable. 

Due to the presence of radioactive decay processes in the crystal, the particles from the radioactive decay propagate through the crystal, resulting from the Cherenkov effect to generate the optical photons ranging from extreme ultraviolet to visible wavelengths, which considering only photon propagation is insufficient. It is also necessary to take into account other decay products, such as electrons, positrons, protons, and $\alpha$ particles. The propagation of these charged particles in the crystal can generate Cherenkov radiation, which may influence the detection of the $^{229m}$Th decay signal. From the calculation, the threshold energies required to produce Cherenkov radiation at a wavelength of 148 nm are approximately 0.15 MeV for electrons, 276 MeV for protons, and 1096 MeV for $\alpha$ particles. For radioactive decays, the energies required for protons and $\alpha$ particles to produce Cherenkov radiation are far above their typical decay energies; therefore, their contribution can be neglected in the detection of the $^{229m}$Th signal. Consequently, only $\beta^{\pm}$ decays need to be considered when evaluating the Cherenkov background, as shown by the energy–time distributions in Fig. 4(b) and Fig. 4(c). As can be seen, almost electron and positrons from $\beta^{\pm}$ decays concerted on above 0.15 MeV, especially the electron contribution, which could contribute to the Cherenkov ground to infer the measurement signal of $^{229m}$Th. Therefore, we depicted the wavelength and emission time distribution of the Cherenkov photons from the radioactive decay of neutron capture reaction residuals of $^{228}$Ra-doped CaF$_2$ crystal in Fig. 4(d). The Cherenkov background exhibits a strong intensity at wavelengths below 140 nm and remains significant within the post-irradiation time window from 10$^6$ to 10$^{10}$ s. In contrast, the radiative decay signal of $^{229m}$Th is concentrated around 148 nm and extends over a time scale from approximately 10$^1$ to 10$^5$s. Despite the presence of the Cherenkov background, a distinct $^{229m}$Th signal can still be identified within the interval from 10$^3$ to 10$^{4.5}$ s, as indicated by the red annotation in Fig. 4(d). Therefore, for the detection of $^{229m}$Th, the measurement time is preferably chosen within 10$^3$ to 10$^{4.5}$s after neutron irradiation. Similar considerations apply to the other two radium-doped crystals, for which the measurement time should also be restricted to this interval.

\begin{figure*}[!htb]
\includegraphics[width=1.0\hsize]{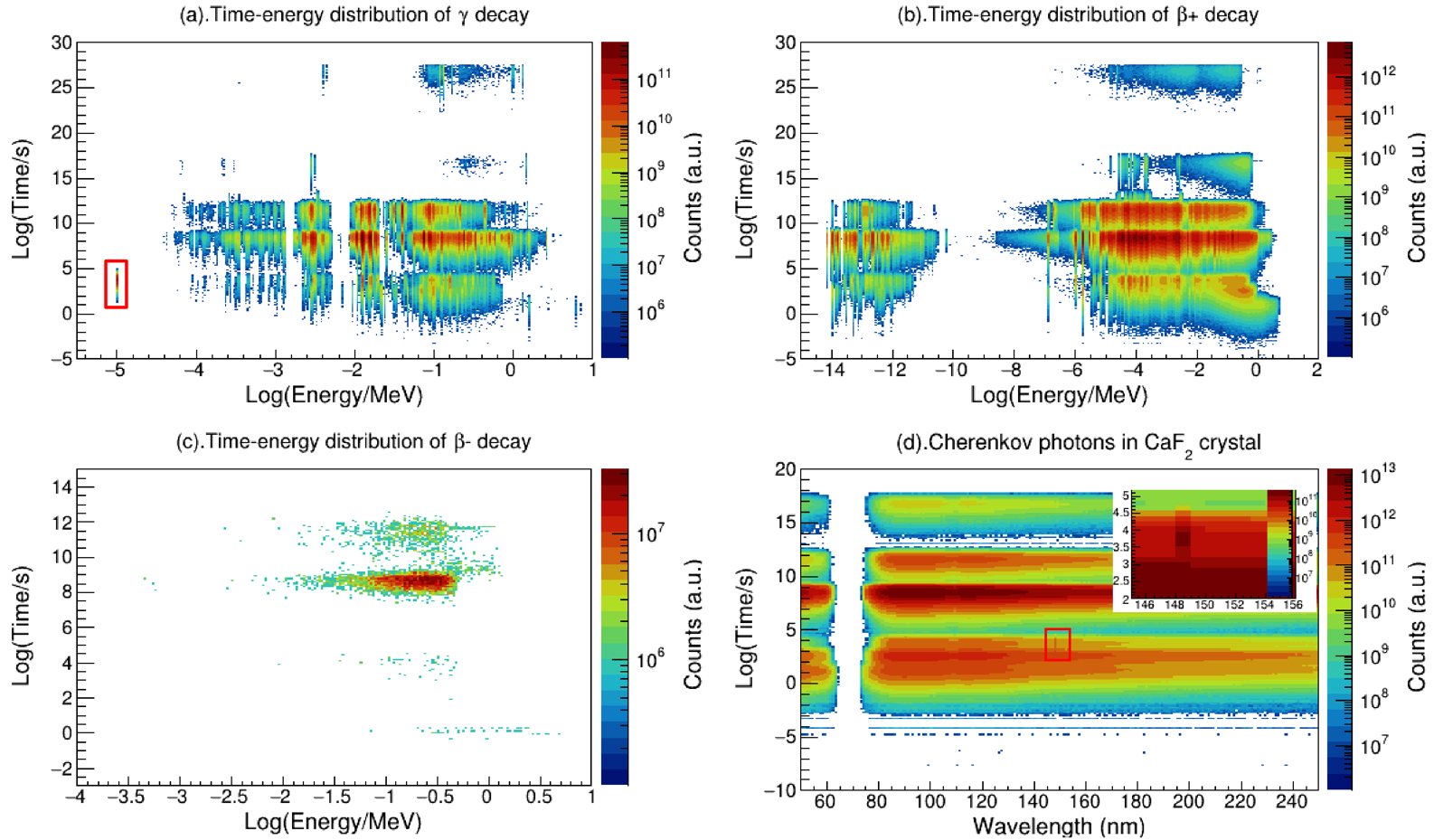}    
\caption{(a) Time-energy distrubtion of $\gamma$ decay emitted from the neutron-irradiated $^{228}$Ra-doped CaF$_2$ crystal. (b) Time-energy distrubtion of $\beta-$ decay emitted from the neutron-irradiated $^{228}$Ra-doped CaF$_2$ crystal. (c) Time-energy distrubtion of $\beta+$ decay emitted from the neutron-irradiated $^{228}$Ra-doped CaF$_2$ crystal. (d) The wavelength-time distribution of the emitted optical photons, involving $^{229m}$Th decay signal and Cherenkov backgrounds. The red box highlights the emitted optical photons around 148 nm associated with the $^{229m}$Th decay signal.}
\end{figure*}

Although the decay signal of $^{229m}$Th can already be identified in the time–wavelength distribution shown in Fig. 4(d), a more direct visualization is provided in Fig. 5, where the wavelength distributions of Cherenkov photons and photons emitted from the decay of $^{229m}$Th are presented for the three radium-doped crystals. As shown in Fig. 5, the radiative signal of $^{229m}$Th around 148 nm remains clearly distinguishable from the Cherenkov background in all three crystals, demonstrating the capability of the proposed method for sensitive detection of the generated $^{229m}$Th.
\begin{figure*}[t]
    \centering
    \includegraphics[width=\textwidth]{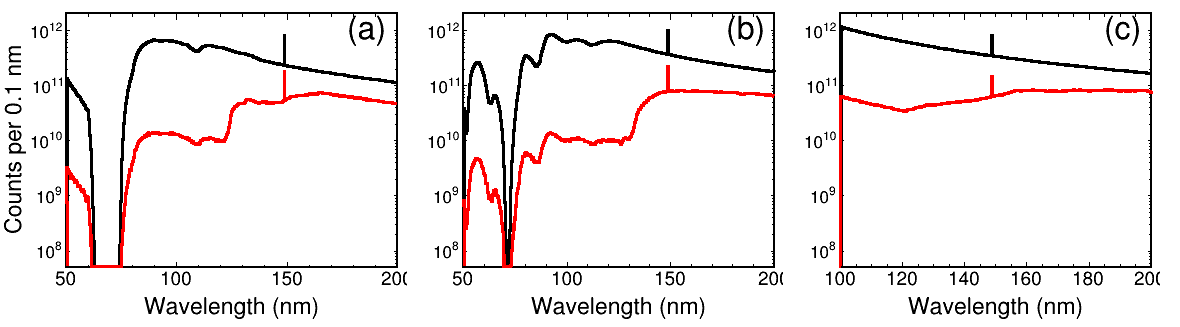}
    \caption{Photon spectra for the three radium-doped crystals integrated over the post-irradiation time interval from $10^3$ to $10^{4.5}$ s. The black curves represent optical photons generated inside the crystals, while the red curves represent optical photons emitted from the crystal surfaces. }  
\end{figure*}

\section{Signal-to-Noise Ratio}
In addition to the radioactive decay products generated following neutron irradiation, the doped $^{228}$Ra itself can also contribute to the background signal in the detection of $^{229m}$Th. Owing to its intrinsic radioactivity and relatively high concentration in the crystal, particles produced in the decay chain of $^{228}$Ra may generate additional Cherenkov radiation and thus interfere with the detection of the $^{229m}$Th signal. Therefore, the background affecting the detection of $^{229m}$Th consists of two main components: the decay signal originating from the intrinsic $^{228}$Ra dopants and the radioactive decay signal from nuclides produced in the crystal following neutron irradiation. To quantify the feasibility of detecting the photon decay signal from $^{229m}$, the signal-to-noise ratio (SNR) of photons emitted from the three crystals was evaluated. The SNR is defined as:

\begin{equation}
\mathrm{SNR}=\frac{S}{\sqrt{B}}
\end{equation}

where S denotes the net counts of the $^{229m}$ signal photons escaping from the crystals, and B represents the background counts originating from the decay of the doped $^{228}$Ra and the radioactive residual nuclei produced by neutron-capture reactions. Figure 6 shows the net counts of the $^{229m}$Th signal photons, the corresponding background noise, and the resulting SNR for photons emitted from the three crystals. The background noise was evaluated within the wavelength range of $148\pm2$ nm, corresponding to the wavelength resolution of most VUV detectors. The signal and background counts were integrated over the post-irradiation time interval from $10^3$ to $10^{3.5}$ s.

\begin{figure}[!htb]
\includegraphics[width=\hsize]{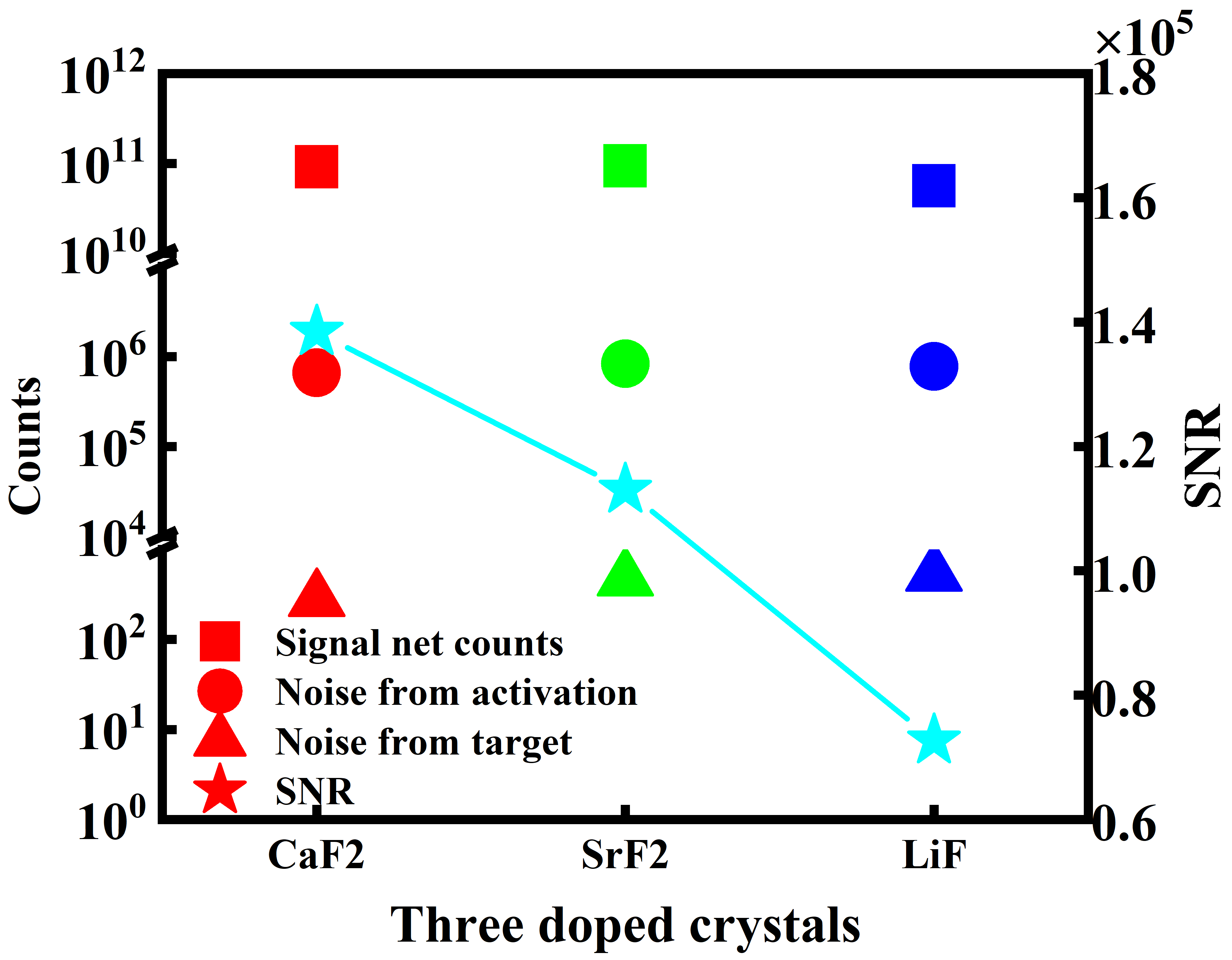}
\caption{Net $^{229m}$Th signal counts, activation-induced noise, target-induced noise, and the corresponding SNR for the three radium-doped crystals. The red, green, and blue symbols correspond to CaF$_2$, SrF$_2$, and LiF, respectively. Squares denote the net counts of the $^{229m}$Th signal, circles represent the activation-induced noise, triangles indicate the noise originating from the intrinsic radioactivity of the doped target, and stars denote the corresponding SNR values. The signal and noise values are evaluated within the post-irradiation time window from $10^3$ to $10^{3.5}$ s.}
\end{figure}

As shown in Fig. 6, the contribution from the intrinsic $^{228}$Ra dopants is relatively small within the selected measurement window of 10$^3$ to 10$^{3.5}$ s. This is mainly due to the long half-life of $^{228}$Ra ($T_{1/2} = 5.75 $  years), which results in a comparatively low decay rate over the time scale relevant for the measurement. Although the radioactive products generated following neutron irradiation produce a considerably larger background, their contribution remains significantly lower than the net $^{229m}$Th signal counts. Consequently, all three radium-doped crystals exhibit high SNRs for the detection of $^{229m}$Th. These results indicate that the decay signal of $^{229m}$Th can be observed with good statistical significance, enabling reliable detection and verification of $^{229m}$Th production. Among the three crystals, CaF$_2$ crystal provides the most favorable conditions for observing and confirming the generation of $^{229m}$Th due to its higher signal yield and superior SNR. Due to the varying spectral resolutions of different detectors, their ability to discriminate photons of nearby wavelengths differs accordingly. To evaluate the impact of detector performance on the observability of the $^{229m}$Th signal, we examine the variation of the background noise and the resulting SNR as a function of the detector wavelength resolution, as shown in Figs. 7(a-c). Since the signal counts of $^{229m}$Th are independent of the detector wavelength resolution, they are taken as constant values of $9.14\times10^{10}$, $9.39\times10^{10}$, and $5.63\times10^{10}$ for CaF$_2$, SrF$_2$, and LiF, respectively, within the time window of 10$^3$ to 10$^{3.5}$ s.

As can be seen from Figs. 7(a-c), improved wavelength resolution leads to a reduction in the background level and consequently an increase in the SNR, thereby enhancing the detectability of the $^{229m}$Th signal. Nevertheless, our results show that even for detectors with relatively poor wavelength resolution, the $^{229m}$Th signal remains clearly distinguishable from the background and retains a high statistical significance. These results indicate that the proposed approach offers a significant advantage for both the production and reliable detection of $^{229m}$Th. 

In addition to the wavelength resolution of the detector, the post-irradiation measurement time also has a significant impact on the detectability of the $^{229m}$Th signal, as indicated in Figs. 4(a) and 4(d). Since $^{229m}$Th is produced through a series of radioactive decay processes, its characteristic emission can only be observed within a limited time interval after irradiation, approximately from $10^1$ to $10^5$ s. A relatively strong $^{229m}$Th signal is mainly concentrated in the time range from $10^{3}$ to $10^{4.5}$ s. However, the optimal detection conditions are determined not only by the intensity of the $^{229m}$Th signal itself, but also by the time-dependent background level within the same measurement window. The background evolves with time due to the decay of radionuclides produced by neutron irradiation, as well as the continuous radioactive decay of the doped $^{228}$Ra in the crystal. Consequently, the SNR may vary significantly for different measurement windows. Consequently, the SNR may vary significantly for different measurement windows. Therefore, the dependences of the net $^{229m}$Th signal counts, the background noise, and the resulting SNR on the post-irradiation measurement time are further evaluated, as shown in Figs. 7(d-f).

\begin{figure*}[t]
    \centering
    \includegraphics[width=0.33\textwidth]{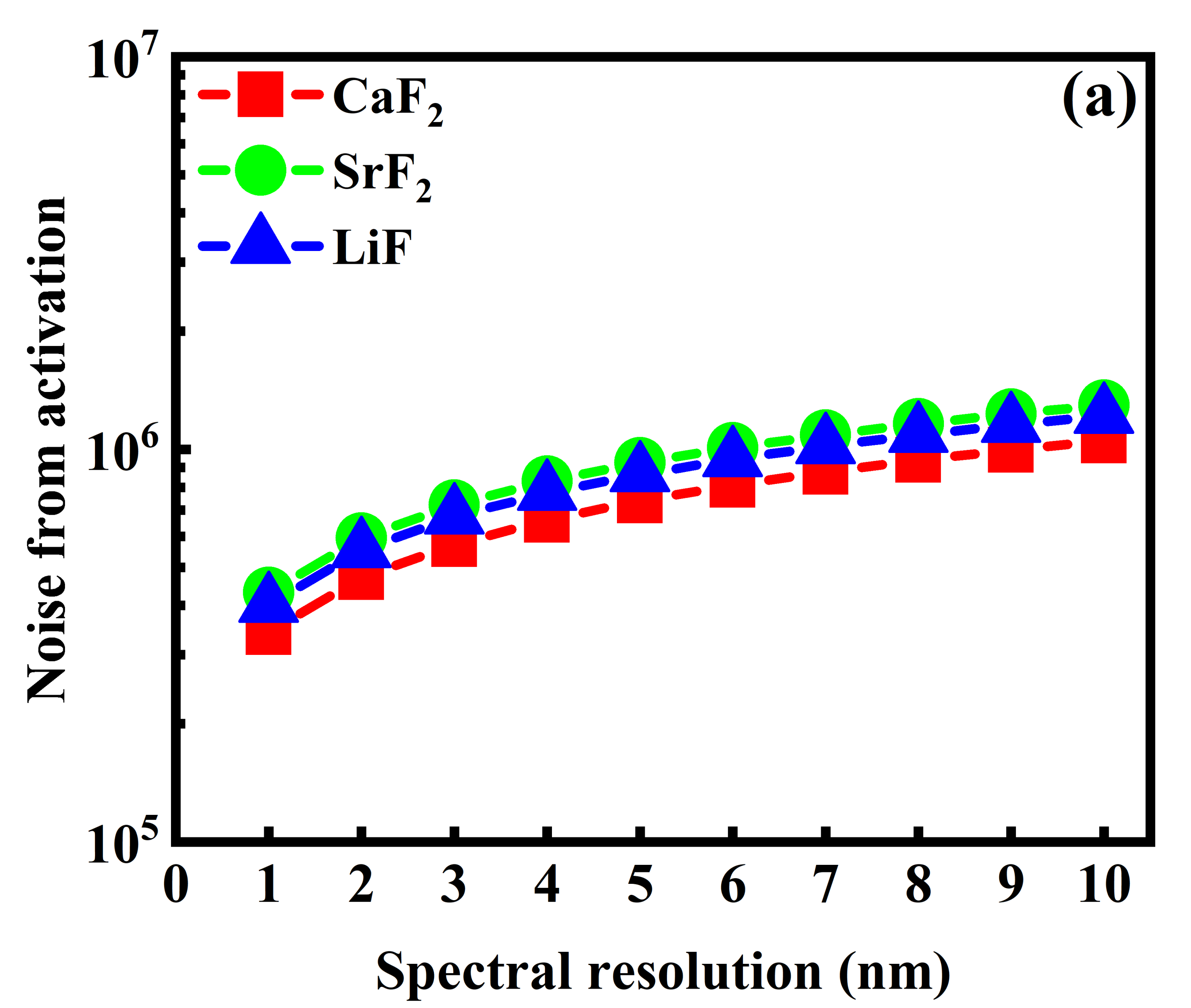}
    \hfill
    \includegraphics[width=0.33\textwidth]{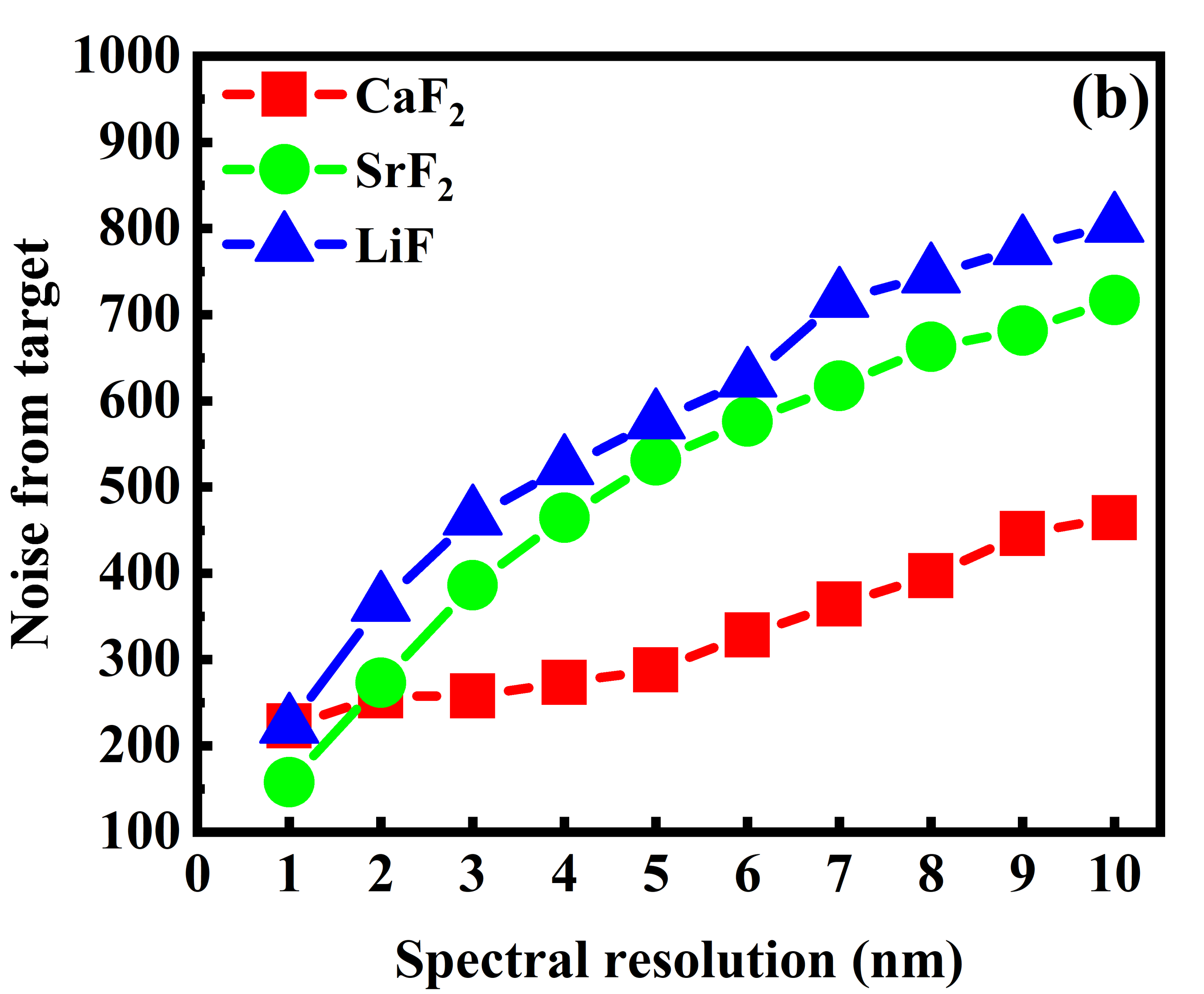}
    \hfill
    \includegraphics[width=0.33\textwidth]{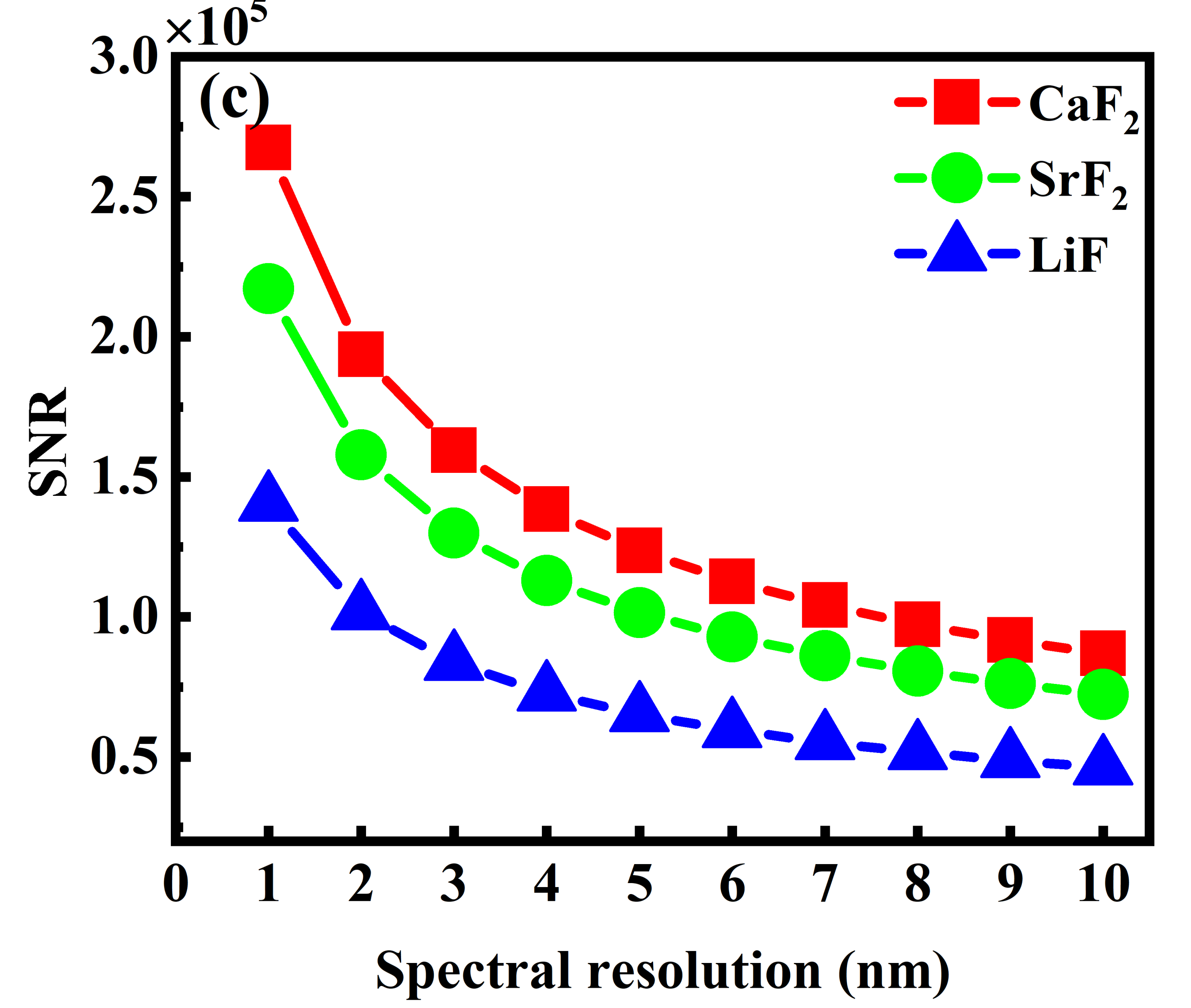}
    \hfill
    \includegraphics[width=0.33\textwidth]{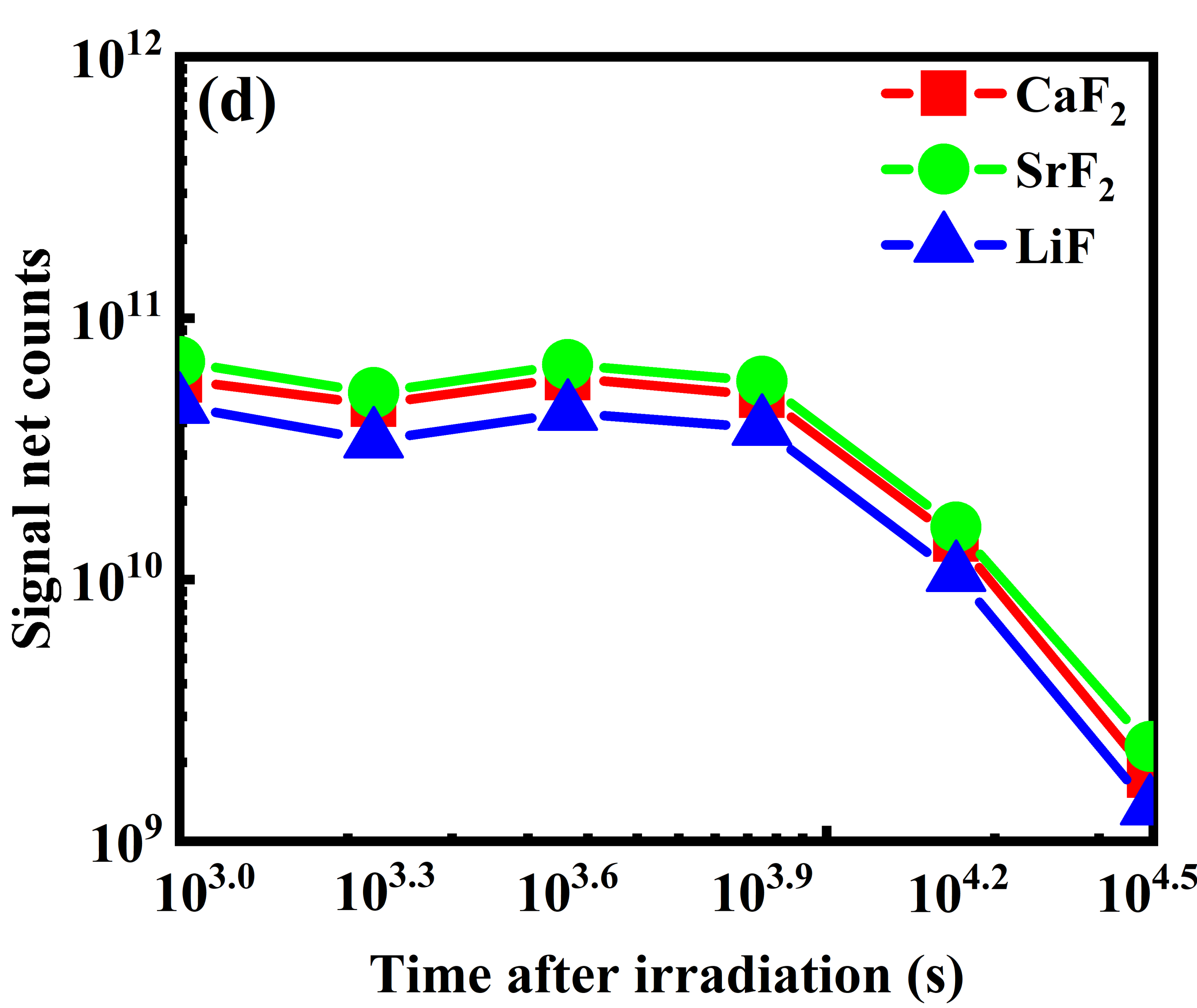}
    \hfill
    \includegraphics[width=0.33\textwidth]{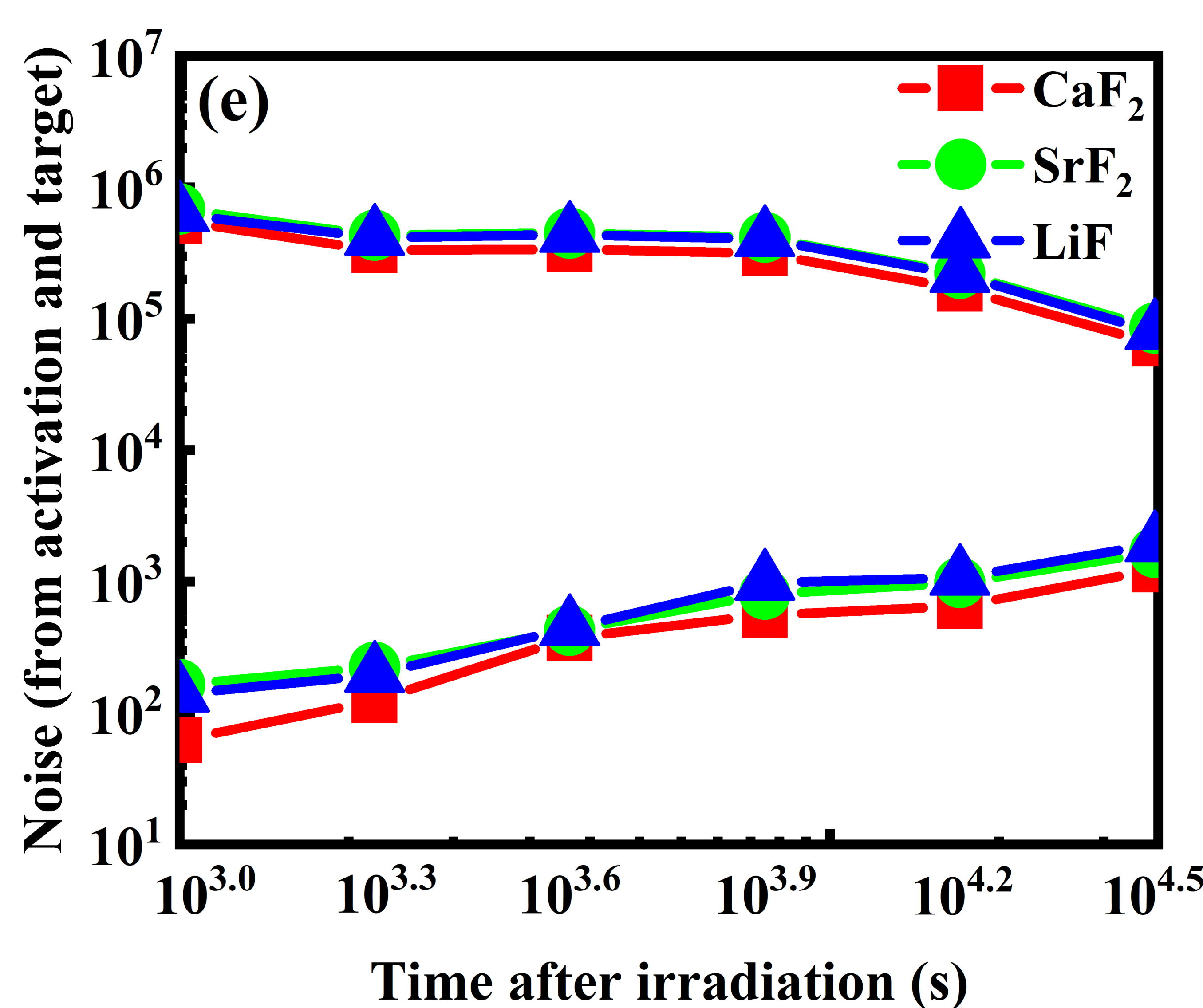}
    \hfill
    \includegraphics[width=0.33\textwidth]{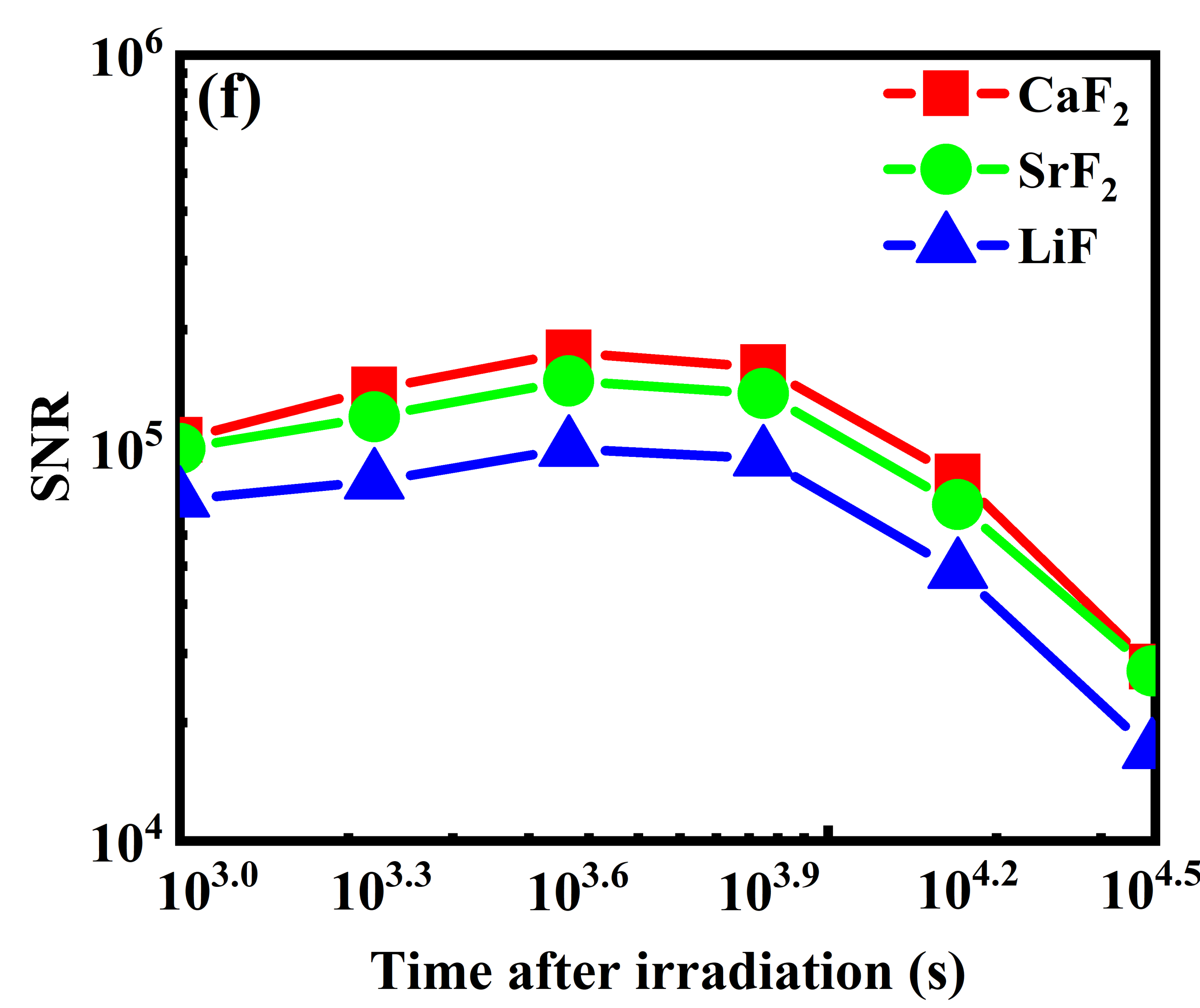}
\caption{Activation-induced noise, target-induced noise, and the resulting SNR as functions of detector wavelength resolution for the three doped crystals, namely CaF$_2$ (a), SrF$_2$ (b), and LiF (c). Net counts of the $^{229m}$Th signal, activation-induced noise, target-induced noise, and the corresponding SNR as functions of post-irradiation measurement time for the three doped crystals, namely CaF$_2$ (d), SrF$_2$ (e), and LiF (f). A fixed counting time of 1000 s is adopted for each measurement window.}
\end{figure*}

As shown in Figs. 7(d-f), the three doped crystals exhibit similar temporal behaviors in the net counts of the $^{229m}$Th signal, the activation background counts, the target background counts, and the resulting SNR following neutron irradiation. In the measurement time window from approximately $10^{3.6}$ s to $10^{3.9}$ s (corresponding to about 4000–8000 s after irradiation), the net counts of the $^{229m}$Th signal approach their maximum values, which is attributed to the decay of $^{229}$Ac, while the activation background counts gradually decrease with time. Although the target background counts originating from the intrinsic decay of the doped $^{228}$Ra increase continuously, their contribution to the total background remains relatively small during this period. As a result, the SNR reaches its maximum within this time interval, making it the optimal measurement window for detecting the $^{229m}$Th signal. At later measurement times, such as around $10^{4.2}$ s (approximately 1.6 × $10^4$ s after irradiation), the $^{229m}$Th signal decreases rapidly due to radioactive decay, whereas the target background counts continue to increase and become a more significant fraction of the total background. Consequently, the SNR gradually decreases, making the detection of the $^{229m}$Th signal less favorable than in the optimal measurement window.

\section{The $^{229}$Th production contribution}
The recent demonstration of continuous-wave absorption spectroscopy of the $^{229}$Th nuclear transition has established a rapid and highly sensitive method for probing the nuclear resonance \cite{bib:52,bib:53,bib:54}. In this technique, the measured signal is governed primarily by the effective optical depth, OD $\propto L\cdot \rho _{Th} \cdot \chi$ (where L is the interaction length, $\rho _{Th}$ the $^{229}$Th concentration, and $\chi$ represents the fraction of nuclei occupying optically addressable defect centers). A quantitative description of the $^{229}$Th concentration profile along the beam direction is therefore essential for accurately predicting the achievable OD and assessing the detection efficiency.

Taking  CaF$_2$ crystal as an example, we perform Geant4 Monte Carlo simulations of the concentration profile of $^{229}$Th produced under the irradiation conditions considered in this work (see Fig. 8). In this simulation, the neutron source is modeled as a spatially uniform surface source incident on the left face of the crystal, with neutrons injected horizontally into the target. The neutron energy spectrum adopted is the one shown in the previous Fig. 2. The crystal is assumed to have a rectangular cuboid geometry with a cross‑section of 1 cm $\times$ 1 cm in the x‑y plane and a length of 10 cm along the z‑axis (the neutron irradiation direction), with a neutron irradiation time of 1 s. 

\begin{figure*}[t]
    \centering
    \includegraphics[width=0.33\textwidth]{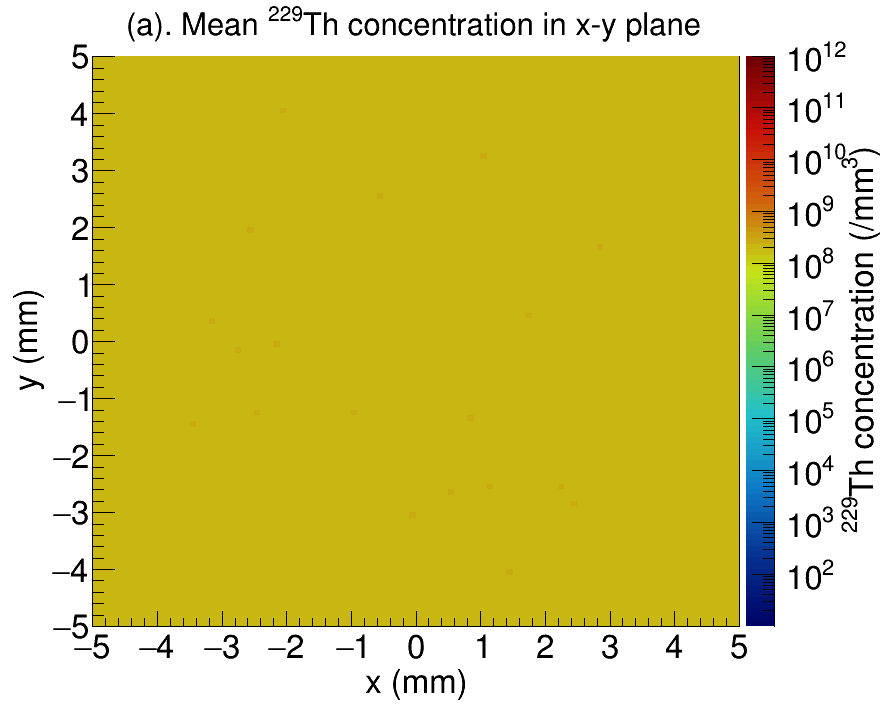}
    \hfill
    \includegraphics[width=0.33\textwidth]{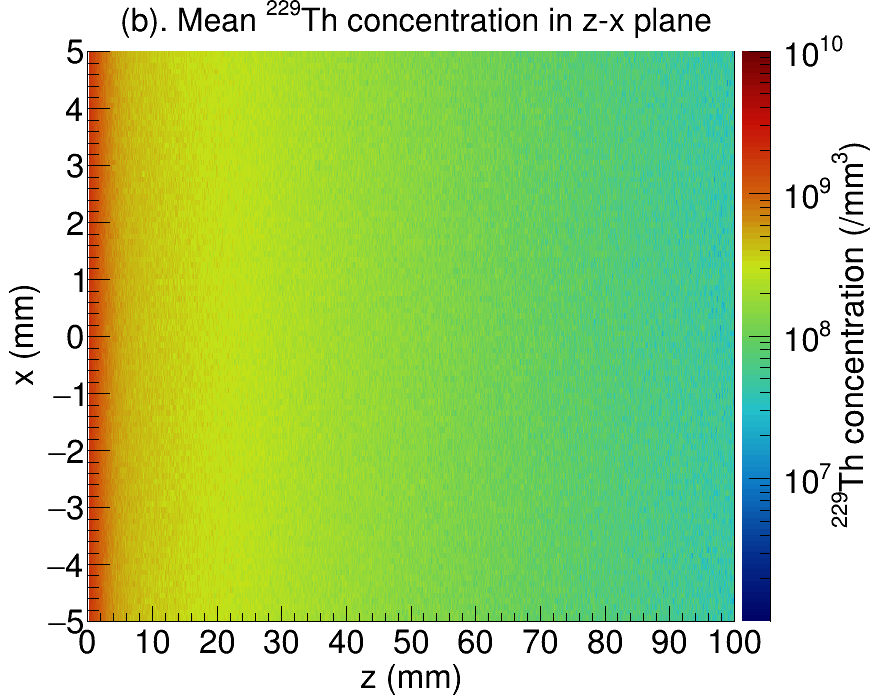}
    \hfill
    \includegraphics[width=0.33\textwidth]{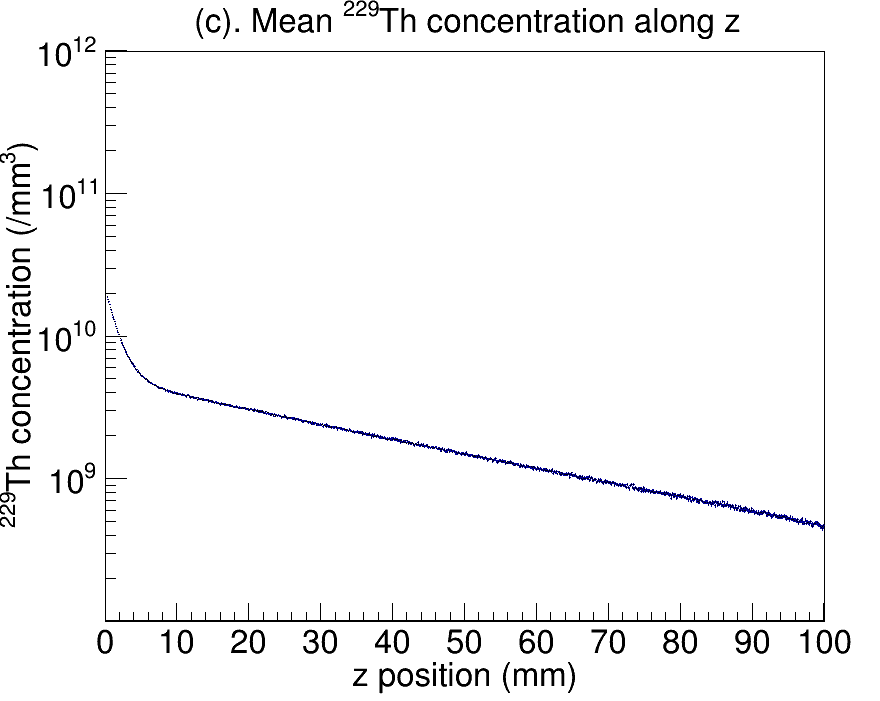}
\caption{Simulated spatial distribution of the $^{229}$Th concentration in the CaF$_2$ crystal. (a) Mean concentration in the x-y plane (averaged over z). (b) Concentration distribution in the z-x plane (averaged over y). (c) Mean concentration along the z direction (averaged over the x-y plane). The bin size is 0.1 mm.}
\end{figure*}

As shown in Fig. 8(a), the $^{229}$Th concentration is nearly uniform across the transverse cross section of the crystal. The local concentration in this region reaches approximately 2.2$\times10^{11}$ cm$^{-3}$ (converted from the simulated value of ~2.2$\times10^{8}$ mm$^{-3}$), demonstrating the feasibility of scaling the production yield toward the concentration range required for future laser excitation experiments. In contrast, Fig. 8(b) and (c) reveal a pronounced reduction in concentration along the depth direction (z‑axis). The concentration decreases from ~1.9$\times10^{12}$ cm$^{-3}$ near the irradiated surface to ~3.0$\times10^{10}$ cm$^{-3}$ at a depth of 10 cm. Notably, the most rapid attenuation occurs within the first centimeter: the concentration drops by nearly half over the initial 6 mm, after which the decline becomes more gradual. 

From the perspective of optimizing the effective optical depth, OD $\propto L\cdot \rho _{Th} \cdot \chi$, these simulation results provide several important implications for continuous-wave VUV absorption spectroscopy. The near-surface $^{229}$Th concentration of approximately 2.2$\times10^{11}$ cm$^{-3}$ represents the maximum achievable density within the irradiated crystal and defines the most favorable region for future resonance-based interrogation schemes. Because the $^{229}$Th concentration decreases rapidly within the first few centimeters owing to neutron attenuation, it is advantageous to employ centimeter-scale crystals and illuminate the sample from the neutron-irradiated surface, thereby maximizing the number of resonant nuclei contributing to the optical absorption signal. 

Beyond optimizing the crystal geometry, the $^{229}$Th concentration can be further increased within the neutron capture scheme. Higher $^{228}$Ra doping concentrations or longer irradiation times in high-flux reactors can proportionally increase the $^{229}$Th yield, providing a practical pathway toward concentrations of $10^{15}-10^{16}$ cm$^{-3}$, which are desirable for solid-state nuclear clock applications. Moreover, under the approximately isotropic neutron fields available in practical reactor irradiation environments, the $^{229}$Th distribution is expected to become substantially more homogeneous throughout the crystal volume than in the one-sided irradiation geometry considered here. 

The choice of host crystal provides an additional degree of flexibility. As summarized in Table 1,  $^{228}$Ra-doped SrF$_2$ crystal produces a considerably higher $^{229}$Th yield than $^{228}$Ra-doped CaF$_2$ crystal, whereas $^{228}$Ra-doped LiF crystal exhibits a lower production yield. These differences arise from the neutron interaction characteristics of the host materials and their achievable $^{228}$Ra doping concentrations, allowing the crystal host to be selected according to specific experimental priorities, such as maximizing the effective optical depth or reducing background radiation.

Overall, the favorable spatial distribution, excellent scalability, improved uniformity achievable under practical irradiation conditions, and the flexibility offered by different host materials establish the neutron-based conversion approach as a promising and versatile route for producing high-quality $^{229}$Th-doped crystals for future solid-state nuclear clock experiments. 

\section{Conclusion}
In this work, we have demonstrated the feasibility of producing and detecting $^{229m}$Th using $^{228}$Ra-doped CaF$_2$, SrF$_2$, and LiF crystals under neutron irradiation. Under a neutron flux of $10^{15}\ \mathrm{n/cm^{2}/s}$ and a $^{228}$Ra doping concentration of $10^{19}\ \mathrm{cm^{-3}}$, this neutron-capture-based approach is capable of producing on the order of $10^{12}$ $^{229}$Th and $^{229m}$Th nuclei within only 1 s of irradiation, demonstrating its potential as a high-yield production route for $^{229m}$Th. In the simulations, both the Cherenkov background induced by radioactive products generated after neutron irradiation and the background arising from the intrinsic decay of the doped $^{228}$Ra were taken into account. Their impact on the detection of $^{229m}$Th was evaluated for the three crystal hosts, namely CaF$_2$, SrF$_2$, and LiF, all of which exhibit high SNRs sufficient for reliable detection of the $^{229m}$Th radiative decay. Furthermore, the effects of detector wavelength resolution and post-irradiation measurement time on the detectability of the $^{229m}$Th signal were systematically investigated. The results indicate that high SNRs can still be maintained with realistic detector wavelength resolutions, and an optimal post-irradiation measurement window was identified for all three crystal hosts. In addition, the spatial distribution of neutron-produced $^{229}$Th within the crystal was systematically investigated. The simulations show that, under one-sided neutron irradiation, the $^{229}$Th yield remains nearly uniform in the transverse direction but decreases with increasing penetration depth because of neutron attenuation. These results suggest that centimeter-scale crystal thicknesses and illumination from the neutron-irradiated surface are favorable for continuous-wave VUV absorption spectroscopy, while a more homogeneous $^{229}$Th distribution is expected under the approximately isotropic neutron fields available in practical high-flux reactor irradiation environments. The methodology developed here also provides a general framework for optimizing neutron-activation production, signal detection, and crystal design in future experimental implementations. Overall, these results provide practical guidance for future experimental studies on neutron-activation production and optical detection of $^{229m}$Th, and support the continued development of solid-state nuclear clocks and precision measurements based on the $^{229m}$Th nuclear transition.

\end{document}